\documentclass[runningheads]{svmult}

\usepackage{makeidx}   
\usepackage{graphicx}  
\usepackage{subeqnar}  
\usepackage{multicol}  
\usepackage{physprbb}  
\makeindex             

\newcommand{\msun}{$M_{\odot}\hspace{1mm}$}
\newcommand{\hs}{\hspace{1mm}}

\newcommand{\nf}{x_{\rm HI}}

\newcommand{\strom}{Str\"omgren sphere}
\newcommand{\stromspace}{Str\"omgren sphere~}
\newcommand{\lya}{Lyman~$\alpha~$}
\newcommand{\lyb}{Lyman~$\beta~$}
\newcommand{\taudamp}{\tau_{D}}
\newcommand{\taures}{\tau_{R}}
\newcommand{\lobs}{\lambda_{\rm obs}}
\newcommand{\zsource}{z_{Q}}

\newcommand{\qname}{SDSS J1030+0524}
\newcommand{\taub}{\tau_{\rm lim(Ly\beta)}}
\newcommand{\taua}{\tau_{\rm lim(Ly\alpha)}}
\newcommand{\tautot}{\tau_{\rm Ly\alpha}}

\def\gsim{\;\rlap{\lower 2.5pt
 \hbox{$\sim$}}\raise 1.5pt\hbox{$>$}\;}
\def\lsim{\;\rlap{\lower 2.5pt
   \hbox{$\sim$}}\raise 1.5pt\hbox{$<$}\;}

\begin{document}
\title*{The Growth of the Earliest Supermassive Black Holes and Their Contribution to Reionization}
\toctitle{The Growth of the Earliest Supermassive Black Holes and Their Contribution to Reionization}
%
%
\titlerunning{The Earliest Supermassive Black Holes}
%
\author{Zolt\'an Haiman \and Mark Dijkstra \and Andrei Mesinger}
\authorrunning{Zolt\'an Haiman et al.}
%
%
\institute{Department of Astronomy, Columbia University, New York, NY 10027, USA}

\maketitle              

\begin{abstract}
We discuss\index{abstract} currently available observational
constraints on the reionization history of the intergalactic medium
(IGM), and the extent to which accreting black holes (BHs) can help
explain these observations.  We show new evidence, based on the
combined statistics of Lyman $\alpha$ and $\beta$ absorption in quasar
spectra, that the IGM contains a significant amount of neutral
hydrogen, and is experiencing rapid ionization at redshift $z\sim
6$. However, we argue that quasar BHs, even faint ones that are below
the detection thresholds of existing optical surveys, are unlikely to
drive the evolution of the neutral fraction around this epoch, because
they would over--produce the present--day soft X--ray background. On
the other hand, the seeds of the $z\sim 6$ quasar BHs likely appeared
at much earlier epochs ($z\sim 20$), and produced hard ionizing
radiation by accretion. These early BHs are promising candidates to
account for the high redshift ($z\sim 15$) ionization implied by the
recent cosmic microwave anisotropy data from WMAP.  Using a model for
the growth of BHs by accretion and mergers in a hierarchical
cosmology, we suggest that the early growth of quasars must include a
super-Eddington growth phase, and that, although not yet optically
identified, the FIRST radio survey may have already detected several
thousand $>10^8{\rm M_\odot}$ BHs at $z>6$.
\end{abstract}

\section{Black Holes and Reionization}

The recent discovery of the Gunn--Peterson (GP) troughs in the spectra
of $z>6$ quasars in the Sloan Digital Sky Survey (SDSS)
\cite{becker01,fan03,white03}, has suggested that the end of the
reionization process occurs at a redshift near $z\sim 6$. On the other
hand, the high electron scattering optical depth, $\tau_e=0.17 \pm
0.04$, measured recently by the {\it Wilkinson Microwave Anisotropy
Probe (WMAP)} experiment \cite{spergel03} suggests that ionizing
sources were abundant at a much higher redshift, $z\sim 15$.  These
data imply that the reionization process is extended and complex, and
is probably driven by more than one population of ionizing sources
(see, e.g., \cite{haiman03} for a recent review).

The exact nature of the ionizing sources remains unknown.  Natural
candidates to account for the onset of reionization at $z\sim 15$ are
massive, metal--free stars that form in the shallow potential wells of
the first collapsed dark matter halos \cite{wl03,cen03,hh03}.  The
completion of reionization at $z\sim 6$ could then be accounted for by
a normal population of less massive stars that form from the
metal--enriched gas in more massive dark matter halos present at
$z\sim 6$.

The most natural alternative cause for reionization is the ionizing
radiation produced by gas accretion onto an early population of black
holes (``miniquasars''; \cite{hl98}).  The ionizing emissivity of the
known population of quasars diminishes rapidly beyond $z\gsim 3$, and
bright quasars are unlikely to contribute significantly to the
ionizing background at $z\gsim 5$ \cite{shapiro94,ham01}.  However, if
low--luminosity, yet undetected miniquasars are present in large
numbers, they could dominate the total ionizing background at $z\sim
6$ \cite{hl98}. Recent work, motivated by the {\it WMAP} results, has
emphasized the potential significant contribution to the ionizing
background at the earliest epochs ($z\sim 15$) from accretion onto the
seeds of would--be supermassive black holes \cite{madau04,ricotti04}.
The soft X--rays emitted by these sources can partially ionize the IGM
early on \cite{oh01,venkatesan01}.

In this contribution, we address the following issues: (1) What is the
fraction of neutral hydrogen in the IGM at $z\sim 6$? (2) Can quasar
black holes contribute to reionization either at $z\sim 6$ or at
$z\sim 15$?  (3) How did BHs grow, in a cosmological context, starting
from early, stellar--mass seeds at $z\sim 20$?  (4) Can we detect
massive early ($z>6$) black holes directly?  Numerical statements
throughout this paper assume a background cosmology with parameters
$\Omega_m=0.27$, $\Omega_{\Lambda}=0.73$, $\Omega_b=0.044$, and
$h=0.71$, consistent with the recent measurements by {\it WMAP}
\cite{spergel03}.

\section{What is the Neutral Fraction of Hydrogen at $z\sim 6$?}

The ionization state of the IGM at redshift $6\lsim z \lsim 7$ has
been a subject of intense study over the past few years.  While the
{\it WMAP} results imply that the IGM is significantly ionized out to
$z\sim 15$, several pieces of evidence suggest that it has a high
neutral fraction at $z\sim 6-7$.

One argument against a simple model, in which the IGM is ionized at
$z\sim 15$, and stays ionized thereafter, comes from the thermal
history of the IGM \cite{hui03,theuns02}.  The temperature of the IGM,
measured from the Ly$\alpha$ forest, is quite high at $z\sim 4$, with
various groups finding values around $T\sim 20,000$K
\cite{zht01,mm01,st00}.  As long as the universe is reionized before
$z = 10$ {\it and remains highly ionized thereafter}, the IGM reaches
an asymptotic thermal state, determined by a competition between
photoionization heating and adiabatic cooling (the latter being due to
the expansion of the universe).  Under reasonable assumptions about
the ionizing spectrum, the IGM then becomes too cold at $z=4$ compared
to observations \cite{hui03}.  Therefore, there must have been
significant (order unity) changes in fractions of neutral hydrogen
and/or helium at $6 < z < 10$, and/or singly ionized helium at $4 < z
< 10$.  An important caveat to this argument is the possible existence
of an additional heating mechanism that could raise the IGM
temperature at $z\sim 4$.  Galactic outflows could heat the IGM, in
principle, but observations of close pairs of lines of sight in lens
systems suggest that the IGM is not turbulent on small-scales, arguing
against significant stir--up of the IGM by winds \cite{rauch01}.  The
known quasar population is likely driving the reionization of helium,
HeII$\rightarrow$HeIII, at $z\sim 3$ (e.g. \cite{heap}). If this
process starts sufficiently early, i.e. at $z\gsim 4$, then HeII
photoionization heating could explain the high IGM temperature.  It
would be interesting to extend the search for HeII patches that do not
correlate with HI absorption to $z\sim 4$ to test this hypothesis.

A second argument for a large neutral fraction comes from the rapid
redshift--evolution of the transmission near the redshifted Ly$\alpha$
wavelength in the spectra of distant quasars.  The Sloan Digital Sky
Survey (SDSS) has detected large regions with no observable flux in
the spectra of several z $\sim$ 6 quasars
\cite{becker01,fan03,white03}.  The presence of these Gunn-Peterson
(GP) troughs by itself only sets a lower limit on the volume weighted
hydrogen neutral fraction of $\nf \gsim 10^{-3}$ \cite{fan02}.
However, this strong limit implies a rapid evolution in the ionizing
background, by nearly an order of magnitude in amplitude, from $z=5.5$
to $z\sim 6$ \cite{cm02,fan02,lidz02} (we note that the Lyman $\beta$
region of the spectra, which is needed for this conclusion, is
dismissed in another recent study \cite{songaila04}, which therefore
reaches different conclusions).  Known ionizing populations (quasars
and Lyman break galaxies) do not evolve this rapidly; comparisons with
numerical simulations of cosmological reionization (e.g.
\cite{gnedin04}) suggests that we are, instead, witnessing the end of
the reionization epoch, with the IGM becoming close to fully neutral
at $z\sim 7$.  At this epoch, when discrete HII bubbles percolate, the
mean--free--path of ionizing photons can evolve very rapidly (e.g.
\cite{hl99}) and could explain the steep evolution of the background
flux.

However, perhaps the strongest argument for a large neutral fraction
comes from the presence of the cosmic Str\"omgren spheres surrounding
high--$z$ quasars.  If indeed the intergalactic hydrogen is largely
neutral at $z\sim 6$, then quasars at this redshift should be
surrounded by large ionized (HII) regions, which will strongly modify
their absorption spectra \cite{mr00,ch00}.  Recent work has shown that
the damping wing of absorption by neutral hydrogen outside the HII
region imprints a feature that is statistically measurable in a sample
of $\sim$ 10 bright quasars without any additional assumptions
\cite{mhc04}.  A single quasar spectrum suffices if the size of the
Str\"omgren sphere is constrained independently \cite{mhc04,mh04}.  In
addition, with a modest restriction (lower limit) on the age of the
source, the size of the HII region itself can be used to place
stringent limits on the neutral fraction of the ambient IGM
\cite{wl04}.

\begin{figure}[ht]
\begin{center}
\includegraphics[width=0.8\textwidth]{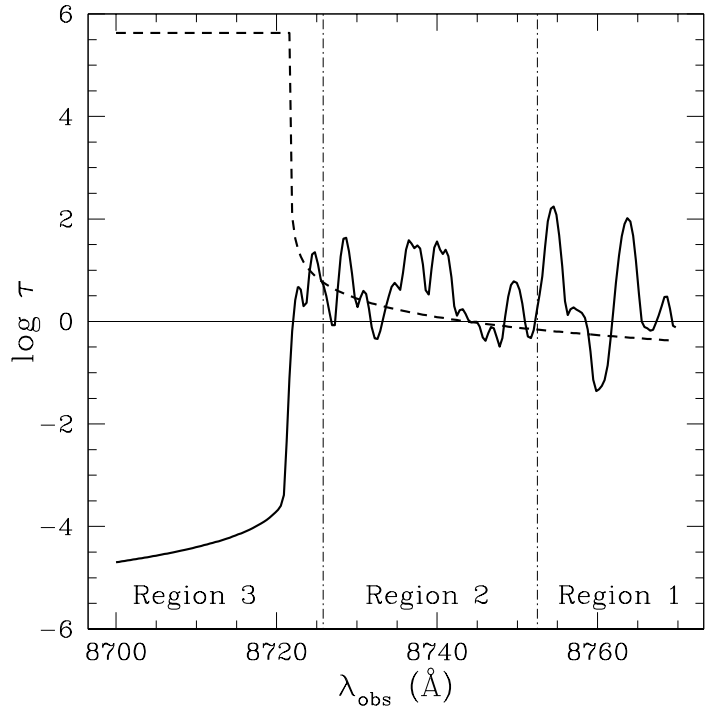}
\end{center}
\vspace{-\baselineskip}
\caption[]{Model from a hydrodynamical simulation for the optical
depth contributions from within ($\tau_R$) and from outside ($\tau_D$)
the local ionized region for a typical line of sight towards a
$\zsource=6.28$ quasar embedded in a fully neutral, smooth IGM, but
surrounded by a local HII region of (comoving) radius $R_S$ = 44 Mpc.
The \emph{dashed curve} corresponds to $\taudamp$, and the \emph{solid
curve} corresponds to $\taures$.  The total \lya optical depth is the
sum of these two contributions, $\tautot$ = $\taures$ + $\taudamp$.
The \emph{dashed-dotted lines} demarcate the three wavelength regions
used for our analysis described in the text.  For reference, the
redshifted \lya wavelength is at 8852 \AA, far to the right off the
plot.}
\label{fig:taus}
\end{figure}

To elaborate on these last arguments, based on the quasar's HII
region, in Figure~\ref{fig:taus} we illustrate a model for the optical
depth to \lya absorption as a function of wavelength towards a
$\zsource=6.28$ quasar, embedded in a neutral medium ($x_{\rm HI}=1$),
but surrounded by a \strom\ with a comoving radius of $R_S = 44$ Mpc.
Around bright quasars, such as those recently discovered
 \cite{fan01,fan03} at $z\sim 6$, the proper radius of such Str\"omgren
spheres is expected to be $R_S$ $\approx$ 7.7 $x_{\rm HI}^{-1/3}$
$(\dot N_Q/ 6.5\times 10^{57}~{\rm s^{-1}})^{1/3}$ $(t_Q/2\times
10^7~{\rm yr})^{1/3}$ $[(1+z_Q)/7.28]^{-1}$ Mpc  \cite{mr00,ch00}. Here
$x_{\rm HI}$ is the volume averaged neutral fraction of hydrogen
outside the \strom\, and $\dot N_Q$, $t_Q$, and $z_Q$ are the quasar's
production rate of ionizing photons, age, and redshift. The fiducial
values are those estimated for the $z=6.28$ quasar J1030+0524
 \cite{hc02,wl04}.  The mock spectrum shown in Figure 1 was created by
computing the \lya opacity from a hydrodynamical simulation (kindly
provided by R. Cen; the analysis procedure is described in
 \cite{mh04}).  The optical depth at a given observed wavelength,
$\lobs$, can be written as the sum of contributions from inside
($\taures$) and outside ($\taudamp$) the \strom, $\tautot = \taures +
\taudamp$.  The residual neutral hydrogen inside the \strom\ at
redshift $z<z_Q$ resonantly attenuates the quasar's flux at
wavelengths around $\lambda_\alpha(1+z)$, where $\lambda_\alpha =
1215.67$ \AA\ is the rest-frame wavelength of the \lya line center. As
a result, $\taures$ is a fluctuating function of wavelength (solid
curve), reflecting the density fluctuations in the surrounding gas.
In contrast, the damping wing of the absorption, $\taudamp$, is a
smooth function (dashed curve), because its value is averaged over
many density fluctuations.  As the figure shows, the damping wing of
the absorption from the neutral universe extends into wavelengths
$\lobs\gsim 8720$ \AA, and can add significantly to the total optical
depth in this region.

\begin{figure}[ht]
\begin{center}
\includegraphics[width=0.8\textwidth]{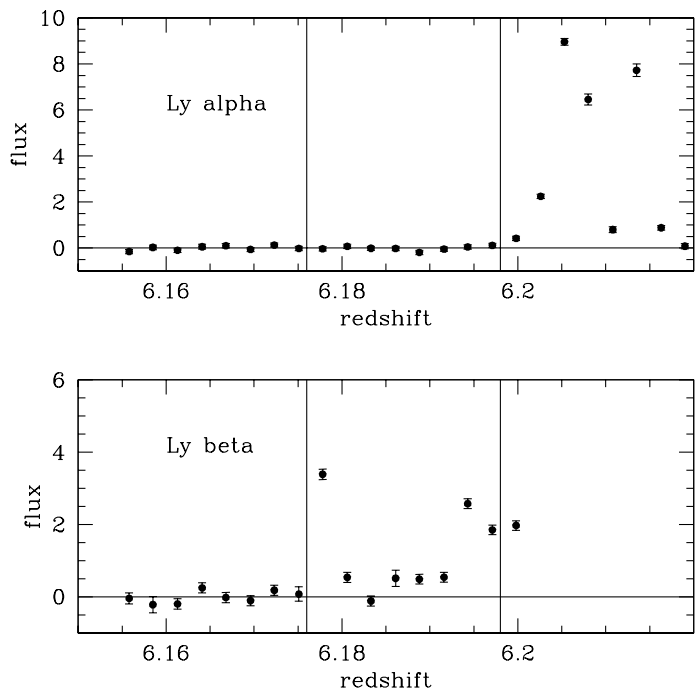}
\end{center}
\vspace{-\baselineskip}
\caption[]{The Keck ESI spectrum of the $z=6.28$ quasar \qname\
\cite{white03}, in units of $10^{-18}~{\rm
erg~s^{-1}~cm^{-2}}$~\AA$^{-1}$, and including uncorrelated $1\sigma$
errors.  The upper and lower panels show the regions of the spectrum
corresponding to Lyman $\alpha$ and $\beta$ absorption in the redshift
range $6.15 < z < 6.28$.  The Lyman $\beta$ cross section is smaller
than Lyman $\alpha$, and a significantly higher column of neutral
hydrogen is required to block the flux in the Lyman $\beta$ region
than in the Lyman $\alpha$ region. This spectrum therefore requires a
steep increase in the opacity over the narrow range from $z=6.20$ to
$z=6.16$, explainable only by a GP damping wing, as shown by the
dashed curve in Figure~\ref{fig:taus}.}
\label{fig:spectrum}
\end{figure}

The sharp rise in $\taudamp$ at wavelengths $\lobs \lsim 8720$ \AA\ is
a unique feature of the boundary of the HII region, and corresponds to
absorption of photons redshifting into resonance outside of the
\strom.  The detection of this feature has been regarded as
challenging: since the quasar's flux is attenuated by a factor of
$\exp(-\tautot)$, an exceedingly large dynamical range is required in
the corresponding flux measurements.  However, simultaneously
considering the measured absorption in two or more hydrogen Lyman
lines can provide the dynamical range required to detect this
feature \cite{mh04}.

In particular, we modeled broad features of the \lya and \lyb regions
of the absorption spectrum of the $z=6.28$ quasar SDSS J1030+0524.
The observational input to our analysis is the deepest available
absorption spectrum of \qname\  \cite{white03}, shown in
Figure~\ref{fig:spectrum}.  The spectrum exhibits a strong \lya
Gunn-Peterson (GP) trough, with no detectable flux between wavelengths
corresponding to redshifts $5.97 < z < 6.20$, as well as a somewhat
narrower \lyb trough between $5.97< z < 6.18$  \cite{white03}, as shown
in Figure~\ref{fig:spectrum}. The flux detection threshold in the \lya
and \lyb regions of this spectrum correspond to \lya optical depths of
$\taua \approx 6.3$ and $\taub \approx 22.8$ respectively
\footnote{For our purposes, these optical depths can be taken as rough
estimates.  Their precise values are difficult to calculate, with
$\taub$ especially uncertain  \cite{songaila04,lidz02,cm02,fan02}.
However, we have verified that our conclusions below remain unchanged
when the threshold opacities are varied well in excess of these
uncertainties. In particular, considering ranges as wide as
$5.5<\taua<7$ and $10<\taub<30$ would lead to constraints similar to
those we derive below.}  

To summarize these constraints, we have divided the spectrum into
three regions, shown in Figure~\ref{fig:taus}.  In Region 1, with
$\lobs \geq$ 8752.5 \AA, the detection of flux corresponds to the {\it
upper} limit on the optical depth $\tautot < 6.3$.  Region 2,
extending from 8725.8 \AA\ $\leq \lobs <$ 8752.5 \AA, is inside the
\lya trough, but outside the \lyb trough. Throughout this region, the
data requires 6.3 $\lsim \tautot \lsim 22.8$.  Region 3, with $\lobs
<$ 8725.8 \AA\, has a {\it lower} limit $\tautot \geq$ 22.8.  As
defined, each of these three regions contains approximately eight
pixels.

We modeled \cite{mh04} the absorption spectrum, attempting to match
these gross observed features. We utilized a hydrodynamical simulation
that describes the density distribution surrounding the source quasar
at $z=6.28$.  We extracted density and velocity information from 100
randomly chosen lines of sight (LOSs) through the simulation box.
Along each line of sight (LOS), we computed the \lya absorption as a
function of wavelength.  The size of the ionized region ($R_S$) and
the fraction of neutral hydrogen outside it ($\nf$), and the quasar's
ionizing luminosity, $L_{\rm ion}$, were free parameters.  Note that
changing $R_S$ moves the dashed ($\taudamp$) curve in
Figure~\ref{fig:taus} left and right, while changing $\nf$ moves it up
and down; changing $L_{\rm ion}$ moves the solid ($\taures$) curve up
and down.  We evaluated $\taures$ and $\taudamp$ for each LOS, and for
each point in a three--dimensional parameter space of $R_S$, $x_{\rm
HI}$, and $L_{\rm ion}$, we computed the fraction of the LOSs that
were acceptable descriptions of the spectrum of \qname, based on the
criteria defined above.  

The procedure outlined above turns out to provide tight constrains on
all three of our free parameters {\it simultaneously}.  In particular,
we find the allowed range for the radius of the \stromspace\ to be 42
Mpc $\leq R_S \leq$ 47 Mpc, and a $\sim 1$ $\sigma$ lower limit on the
neutral fraction of $\nf\gsim 0.17$.  These results can be interpreted
as follows.  As mentioned previously, the presence of flux in Region 2
($\tautot <$ 22.8) sets an immediate {\it lower limit} on
$R_S$. Region 3, however, yields an {\it upper limit} on $R_S$, from
the requirement that $\tautot >$ 22.8 in that region. This high
optical depth cannot be maintained by $\taures$ alone, without
violating the constraint in Region 1 of $\tautot <$ 6.3.  We note that
tight constraints on the neutral fraction ($\nf\gsim 0.1$) can be
obtained from the size of the \stromspace\, together with an assumed
lower limit on its lifetime  \cite{wl04}. Our direct determination of
the \strom\ size is only slightly larger than the value assumed in
 \cite{wl04}, lending further credibility to this conclusion.  

Our direct constraint on the neutral hydrogen fraction, $\nf$, on the
other hand, comes from the presence of flux in the \lyb region of the
spectrum corresponding to Region 2. Because of fluctuations in the
density field (and hence in $\taures$), a strong damping wing is
needed to raise $\tautot$ above 6.3 throughout Region 2, {\it while
still preserving} $\tautot < 6.3$ in Region 1.  This result is derived
from the observed sharpness of the boundary of the HII region alone,
and relies only on the gross density fluctuation statistics from the
numerical simulation.  {\it In particular, it does not rely on any
assumption about the mechanism for the growth of the HII region.}

Using a large sample of quasars (and/or a sample of gamma--ray burst
afterglows with near-IR spectra) at $z>6$, it will be possible to use
the method presented here to locate sharp features in the absorption
spectrum from intervening HII regions, not associated with the target
source itself.  The Universe must have gone through a transition epoch
when HII regions, driven into the IGM by quasars and galaxies,
partially percolated and filled a significant fraction of the volume.
The detection of the associated sharp features in future quasar
absorption spectra will provide a direct probe of the the 3D topology
of ionized regions during this crucial transition epoch (in
particular, it should enhance any constraint available from either the
Lyman $\alpha$ or $\beta$ region alone \cite{furl04}).

\section{Did Accreting Black Holes Contribute to Reionization?} 

The two most natural types of UV sources that could have reionized the
IGM are stars or accreting black holes.  Deciding which of these two
sources dominated the ionization has been studied for over 30 years
(e.g.  \cite{aw72}).  It has become increasingly clear over the past
decade that the ionizing emissivity of the known population of bright
quasars diminishes rapidly beyond $z\gsim 3$, and they are unlikely to
contribute significantly to the ionizing background at $z\gsim 5$
\cite{shapiro94,ham01}.  This, however, leaves open two possibilities.
First, if low--luminosity, yet undetected miniquasars are present in
large numbers, they could still dominate the total ionizing background
at $z\sim 6$ \cite{hl98}.  Second, the supermassive black holes at
$z\sim 6$ must be assembled from lower--mass seeds which accrete and
merge during the hierarchical growth of structure.  The population of
accreting seed BHs can contribute to the ionization of the IGM at
$z\sim 20$ \cite{madau04,ricotti04}.

\begin{figure}[ht]
\begin{center}
\includegraphics[width=0.8\textwidth]{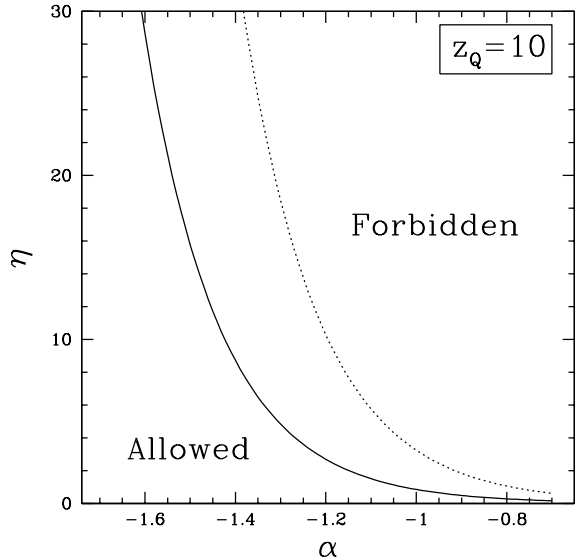}
\end{center}
\vspace{-5\baselineskip}
\caption[]{Constraints on the number of ionizing photons per H atom,
$\eta$, and the power--law index for the slope of the ionizing
background, $\alpha$, based on the intensity of the present--day
SXB. The quasars are assumed to form at $z_Q=10$ and have a power--law
spectrum, $F_E \propto E^{-\alpha}$ for $E> 13.6$ eV.  The curves
bracket the allowed parameter space for the mean (solid line) or
maximum (dotted) unaccounted flux in the SXB at $\sim$1 keV.  The
curves shift by $+0.1$ in $\alpha$ if $z_Q=15$ is assumed.  }
\label{fig:xrb}
\end{figure}

The above two possibilities can both involve faint BHs that are not
individually detectable. However, a population of acceting BHs at
$z\gsim 6$ would be accompanied by the presence of an early X-ray
background.  Since the IGM is optically thick to photons with energies
$E$ below $E_{\rm max} = 1.8 [(1+z)/15)]^{0.5} x_{\rm HI}^{1/3}\hs
{\rm keV}$, the soft X-rays with $E\lsim E_{\rm max}$ would be
consumed by neutral hydrogen atoms and contribute to reionization.
However, the background of harder X-rays would redshift without
absorption and would be observed as a present--day soft X--ray
background (SXB).  Under the hypothesis that accreting BHs are the
main producers of reionizing photons in the high--redshift universe,
it is a relatively straightforward exercise to calculate their
contribution to the present--day SXB.

We assumed for simplicity \cite{dhl04} that the accreting BHs form in
a sudden burst at redshift $z=z_Q$.  The total number of BHs was
expressed by a normalization constant $\eta$, defined as the ratio of
the total number of ionizing photons emitted per unit volume produced
by the BH population to the number density of hydrogen atoms.  Full
ionization of the IGM at $z\sim 6$, where recombinations are
significant, likely requires $\eta\gsim 10$ \cite{ham01}. We find that
in order to account for the electron scattering optical depth
$\tau_e\sim 0.17$ by partially (pre--)ionizing the IGM at $z\sim 15$,
a somewhat smaller $\eta$ is sufficient (see below).

The spectrum of the ionizing background is a crucial ingredient of the
modeling, and depends on the type of accreting BH that is considered.
For luminous quasars powered by supermassive black holes, we adopted a
composite spectrum \cite{saz04}, based on the populations of lower
redshifts ($0< z < 5$) QSOs.  The spectra of lower--mass miniquasars,
with BHs whose masses are in the range $M_{\rm bh}\approx 10^{2-4}$
\msun, are likely to be harder.  For these sources, we followed
\cite{madau04} and adopted a two--component template that includes a
multi--temperature accretion disk, and a simple power-law emission to
mimic a combination of Bremsstrahlung, synchrotron, and inverse
Compton emission by a non-thermal population of electrons.

Finally, we took the unresolved soft X-ray band in the energy range
$0.5-2.0$ keV to be in the range $(0.35-1.23)\times10^{-12}~{\rm
erg~s^{-1}~cm^{-2}~deg^{-2}}$.  This range was obtained  \cite{dhl04}
from a census of resolved X--ray sources, which we subtracted from the
total SXB.  We included the uncertainties in both the measurement of
the total SXB (which was dominant) and in the number of resolved point
sources.  We examined a range of X--ray energies, and found the
strongest constraints in the $0.5-2$ keV band.  We note that a recent
study \cite{bauer04} of faint X--ray sources found a significant
population of star--forming galaxies among these sources, with a
steeply increasing fractional abundance (over AGNs) toward low X--ray
luminosities.  If this trend continues to a flux limit that is only
modestly below the current point--source detection threshold in the
deepest Chandra fields, then the SXB would be saturated, strengthening
our constraints (leaving less room for any additional, high--$z$
quasars).

We find that models in which $z>6$ accreting BHs alone fully reionize
the universe saturate the unresolved X--ray background at the $\geq
2\sigma$ level. Pre-ionization by miniquasars requires fewer ionizing
photons, because only a fraction of the hydrogen atoms need to be
ionized, the hard X--rays can produce multiple secondary ionizations,
and the clumping factor is expected to be significantly smaller than
in the UV--ionization case \cite{ham01,oh01}.  We find that models in
which X--rays are assumed to partially ionize the IGM up to $x_e \sim
0.5$ at $6\lsim z\lsim 20$ are still allowed, but could be constrained
by improved future determinations of the unresolved component of the
SXB.

As emphasized above, the spectral shape of the putative typical
high--$z$ accreting BH is uncertain; the existing templates, motivated
by lower--redshift sources, can be considered merely as guides. Figure
\ref{fig:xrb} shows which combinations of $\alpha$ (the logarithmic
slope of the ionizing spectrum) and $\eta$ are allowed by the
unaccounted flux in the SXB (the solid and dotted curves cover our
inferred range of the unresolved SXB).  This figure shows that for
$\eta=10$, a power--law shallower than $\alpha \approx 1.2-1.4$ will
saturate the unaccounted flux.  For comparison, $\alpha$ is in the
range$-1.5 \lsim \alpha \lsim -0.5$ for $z \lsim 0.3$, and $-1.2 \lsim
\alpha \lsim -0.6$ for $1 \lsim z\lsim 6$ for optically selected radio
quiet quasars \cite{vignali03}.

Our constraints derive from the total number of ionizing photons that
the population as a whole needs to produce to either fully or
partially reionize the universe. Therefore, our conclusions depend
mostly on the assumed spectral shape and the required number of
ionizing photons per hydrogen atom $\eta$.  They are independent of
the details of the population, such as the luminosity function and its
evolution with redshift. Future improvements in resolving the SXB,
improving the limits on the unresolved component by a factor of a few,
would place stringent constraints on the contribution of $z\sim 15$
accreting BHs to the scattering optical depth measured by WMAP.

\section{How Did Black Holes Grow by Accretion and Mergers?} 

Since the seeds of early BHs may have played a role in reionization,
it is all the more interesting to ask how the earliest BH population
was formed, and how it evolved.  The remnants of metal--free
population III stars that form in the first collapsed dark halos can
serve as the initial $\sim 100$\msun seeds that later accrete and
merge together, to give rise to the supermassive BHs making up the
quasar population at $0<z<6$, and the remnant BHs found at the centers
of local galaxies.  Several recent studies
 \cite{kh00,mhn01,vhm03,its03} have addressed various aspects of the
evolution of the BH populations, using the underlying merger trees of
dark matter halos.

Recent work on the generation of gravitational waves during the
coalescence of a binary black hole  \cite{fhh04,merritt04} has
suggested that the binary experiences a typical gravitational recoil
velocity that may be as large as $\gsim 100~{\rm km~s^{-1}}$.  These
velocities exceed the escape velocity $v_{\rm esc}$ from typical dark
matter halos at high--redshift ($z\gsim 6$), and can therefore disrupt
the early stages of growth of BHs by ejecting the earliest seeds from
their host galaxies.  BHs can then start growing effectively only once
the typical dark matter potential wells are sufficiently deep to
retain the recoiling BHs.

Relatively little time is available for the growth of
few$\times10^9~{\rm M_\odot}$ SMBHs prior to $z\sim 6$, and their seed
BHs must be present as early as $z\sim 10$  \cite{hl01}.  A model in
which stellar seed BHs appear in small progenitor DM halos is
consistent with the presence of a $\sim 4\times10^9~{\rm M_\odot}$
SMBH at $z\sim 10$, provided that each seed BH can grow at least at
the Eddington--limited exponential rate, and that the progenitor halos
can form seed BHs sufficiently early on  \cite{hl01}, in halos with
velocity dispersions of $\sigma\sim 30$ km/s.

We quantified the effect of gravitational recoil on the growth of a
few $\times10^9~{\rm M_\odot}$ black hole \cite{kicks}, by assuming
that progenitor holes are ejected from DM halos with velocity
dispersions $\sigma<v_{\rm kick}/2$, and do not contribute to the
final BH mass. Each halo more massive than this threshold was assumed
to host a seed BH that accretes at the Eddington rate, and the BHs
were assumed to coalesce when their parents halos merged.  We took, as
an example, the SMBH powering the most distant SDSS quasar, SDSS
1054+1024 at redshift $z=6.43$, with an inferred BH mass of $\sim
4\times 10^9~{\rm M_\odot}$.  We find that recoil velocities with
$v_{\rm kick}\gsim 65~{\rm km~s^{-1}}$ must occur infrequently, or
else this SMBH must have had a phase during which it gained mass
significantly more rapidly than an Eddington--limited exponential
growth rate (with a radiative efficiency of $\sim 10\%$) would imply.
The super--Eddington growth phase can be avoided \cite{ym04} if seed
BHs can form and grow by mergers in dark halos with a velocity
dispersion as small as $\sigma\approx 5$ km/s, the halos have steep
density profiles (increasing the escape velocity \cite{mq04} from the
central regions by a factor of $\gsim 5$ relative to the naive formula
$v_{\rm esc}=2\sigma$), and the BHs that are retained during the
mergers of their halos can grow uninterrupted at the Eddington rate
between their birth and $z\approx 6$.

\section{Can We Detect Massive $z>6$ BHs Directly?} 

A natural question to ask is whether massive BHs at $z>6$ can be
directly detected.  While the SDSS has detected a handful of
exceptionally bright, few $\times10^9~{\rm M_\odot}$ black holes (the
BH mass is inferred assuming these sources shine at the Eddington
luminosity), they are likely a ``tip of the iceberg'', corresponding
to the rare massive tail of the BH mass function.  In X--rays bands,
the deepest Chandra fields have reached the sensitivity to detect
nearly $\sim 100$ times smaller holes ( \cite{hl99b}, provided they
radiate the Eddington luminosity, with a few percent of their emission
in the X--ray bands).  However, due to the small size of these fields,
they have revealed only a handful of plausible candidates
 \cite{aetal01,barger03}.

Detections however, seem promising in the large radio survey, FIRST.
We used a physically motivated semi--analytic model, based on the mass
function of dark matter halos, to predict the number of radio--loud
quasars as a function of redshift and luminosity \cite{hqb04}.  Simple
models, in which the central BH mass scales with the velocity
dispersion of its host halo as $M_{\rm bh}\propto \sigma_{\rm halo}^5$
have been previously found to be consistent with a number of
observations, including the optical and X--ray quasar luminosity
functions \cite{hl98,wl03b}.  We find that similar models, when
augmented with an empirical prescription for the radio loudness
distribution, overpredict the number of faint ($\sim 10\mu$Jy) radio
sources by 1--2 orders of magnitude. This translates into a more
stringent constraint on the low--mass end of the quasar black hole
mass function than is available from the Hubble and Chandra Deep
Fields. We interpret this discrepancy as evidence that black holes
with masses $\lsim 10^7~{\rm M_\odot}$ are either rare or are not as
radio-loud as their more massive counterparts.  Models that exclude
BHs with masses below $10^7~{\rm M_\odot}$ are in agreement with the
deepest existing radio observations, but still produce a significant
tail of high--redshift objects.  In the 1-10GHz bands, at the
sensitivity of $\sim 10\mu$Jy, we find surface densities of $\sim
100$, $\sim 10$, and $\sim 0.3$ deg$^{-2}$ for sources located at
$z>6$, $10$, and $15$, respectively.  The discovery of these sources
with instruments such as the {\em Allen Telescope Array (ATA)}, {\em
Extended Very Large Array (EVLA)}, and the {\em Square Kilometer Array
(SKA)} would open a new window for the study of supermassive BHs at
high redshift.  We also find surface densities of $\sim 0.1$
deg$^{-2}$ at $z > 6$ for mJy sources that can be used to study 21 cm
absorption from the epoch of reionization.  These models suggest that,
although not yet optically identified, the FIRST survey may have
already detected several thousand such $>10^8{\rm M_\odot}$ BHs at
$z>6$.

\vspace{\baselineskip} ZH thanks the organizers of the conference for
their kind invitation.  We also thank our recent collaborators,
Geoffrey Bower, Renyue Cen, Lam Hui, Avi Loeb and Eliot Quataert for
many fruitful discussions, and Rick White for an electronic version of
the spectrum of \qname.  ZH acknowledges financial support by NSF
through grants AST-0307200 and AST-0307291 and by NASA through grant
NAG5-26029.

\end{document}